\documentclass[a4paper]{article}
\usepackage{icrc2013}
\usepackage[english]{babel}
\title{Impact of the Cherenkov Telescope
Array (CTA) altitude on Dark Matter searches in the Milky Way
Halo}

\authors{Louise Oakes$^1$, Emrah Birsin$^1$, Gernot Maier$^2$, Ullrich
 Schwanke$^1$\\for the CTA Consortium
}

\afiliations{
$^1$ Humboldt Universit\"at zu Berlin \\
$^2$ DESY, Zeuthen \\
}

\email{loakes@physik.hu-berlin.de}

\abstract{Observations of dwarf galaxies and of the Milky Way halo with current
ground-based Cherenkov telescopes have resulted in interesting limits on the
cross-section for dark matter (DM) self-annihilation for WIMP masses above
some 100 GeV. The future Cherenkov Telescope Array (CTA) is expected to
further explore the parameter space of dark matter candidates that are
predicted in extensions of the standard model of particle physics. Due to
its low energy threshold (of order of few tens of GeV) and high sensitivity, CTA will also probe lower WIMP masses than current experiments, but the actual performance in this regime will be influenced by the altitude of the observatory above sea level.
Using the response of possible CTA candidate arrays
to simulated photons and hadrons, we estimate how searches for a WIMP
annihilation signal from the Milky Way halo will be influenced by
altitude of different possible CTA sites.
}

\begin{document}
\maketitle

\section{Introduction}
The Cherenkov Telescope Array (CTA) is a future gamma-ray observatory which is currently in the design and planning stage. An important planning decision is the location of the final sites for building the observatory, for which many factors are under consideration. 
A comprehensive study of the DM search reach of CTA was carried out in~\cite{bib:apdm}; this work extends the studies of the Galactic Centre region to include the effect of the different altitudes of candidate CTA sites, and adds an alternative analysis technique for background subtraction.

CTA will consist of telescopes of three sizes:  Small Size Telescopes (SSTs) sensitive in the highest photon energy ($>10$ TeV) range,  Medium Size Telescopes (MSTs) of 10-12m diameter for the core energy range of 100 GeV to a few TeV and Large Size Telescopes (LSTs) with a diameter of 23m, sensitive to the lowest energy photons. Another important consideration for the performance of CTA is the number and layout of each type of telescope. Several arrays are currently under analysis; two possible array layouts for the Southern Hemisphere site are used in this study: array E (optimised for good performance across the whole energy range) and array B (optimised for best performance at low energies). A site in the Northern Hemisphere is also planned, however the Galactic Centre is better visible in the Southern Hemisphere, therefore the possible Northern site is not included at this stage in the analysis.

\section{Galactic Centre and Halo \label{sec_bg}}
The Galactic Centre is a promising target for indirect DM searches, due to the expected high density of DM in this region. N-body simulations~\cite{bib:Nbody} indicate that the annihilation signal from DM particles, for an observer within the Milky Way, comes dominantly from diffuse DM in the main Galactic Halo (rather than smaller sub-halos).
However, the presence of the Galactic Centre source, HESS J1745-290, and diffuse emission from the Galactic Plane significantly complicate DM searches in this region. 

Radial DM density profiles for Milky Way-like galaxies have been obtained in N-body simulations such as Aquarius~\cite{bib:Nbody} and Via Lactea II~\cite{bib:Nbody2}. Parameterisations of these profiles show large differences close to the the centre of the Milky Way halo, but they agree within a factor of 2 at distances of greater than 10 pc. Regions slightly outside of the Galactic Plane are thus more suitable for DM searches, avoiding the strong GC astrophysical emission and yet still provide a significant gamma-ray flux from DM annihilation. A choice of angular distance of 0.3$^{\circ}$ from the GC is made based on the angular resolution of Cherenkov Telescopes and knowledge of the diffuse emission from the Galactic Plane~\cite{bib:apdm}.

\begin{figure}[ht]
  \centering
  \includegraphics[width=0.4\textwidth]{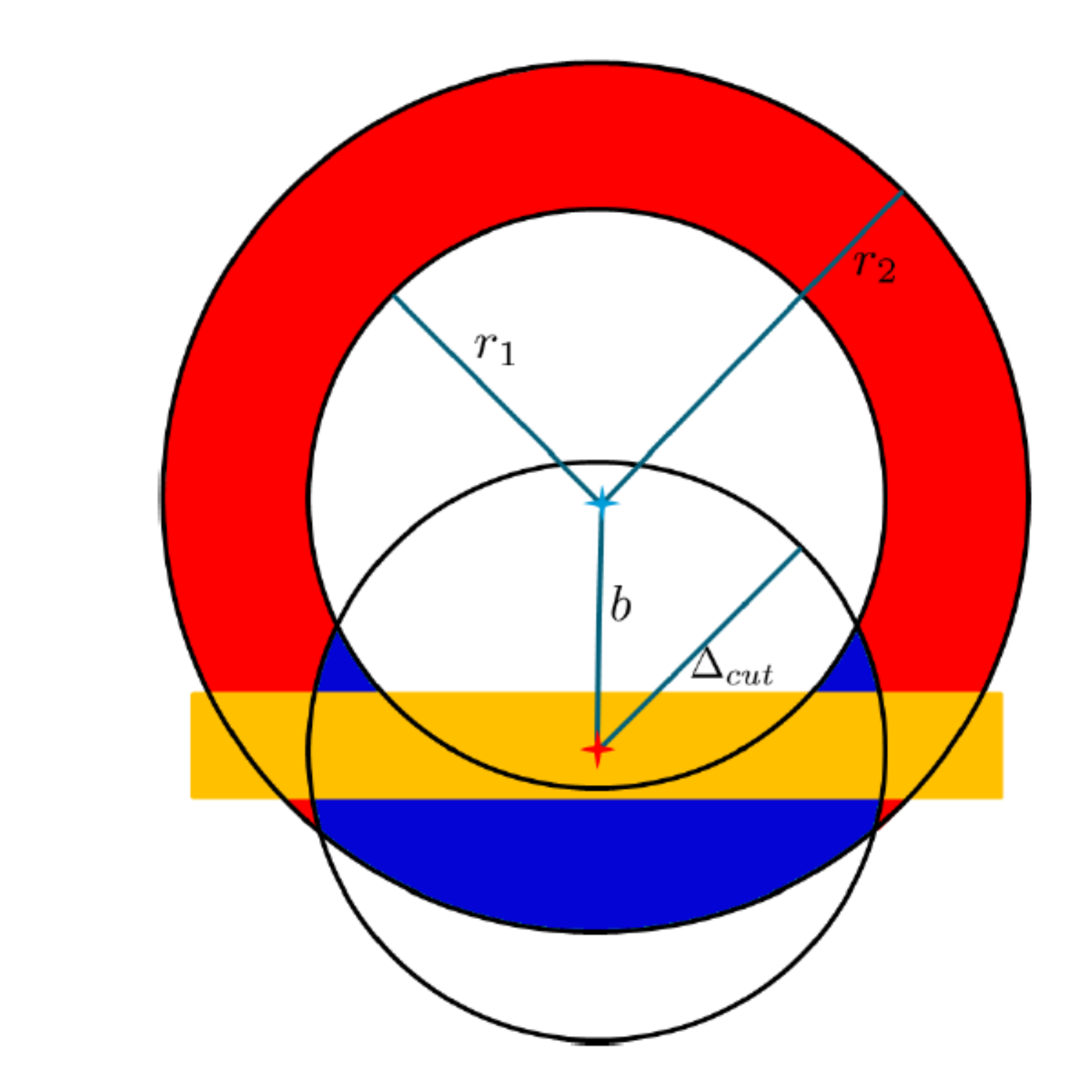}
  \caption{Diagram showing the Ring Background method~\cite{bib:apdm}. Signal and background regions are constructed within a ring of inner radius r1 and outer radius r2, of equal acceptance around the observation position (blue star). The signal region (blue) is defined by the intersection of this ring with a circle of radius $\delta_{cut}$ around the Galactic Centre. The remaining red region of the ring is the background region. A range of $\pm0.3^{\circ}$ around the galactic plane is excluded (shown in yellow).}
  \label{fig1}
 \end{figure}

\begin{figure}[htb]
  \centering
  \includegraphics[width=0.4\textwidth]{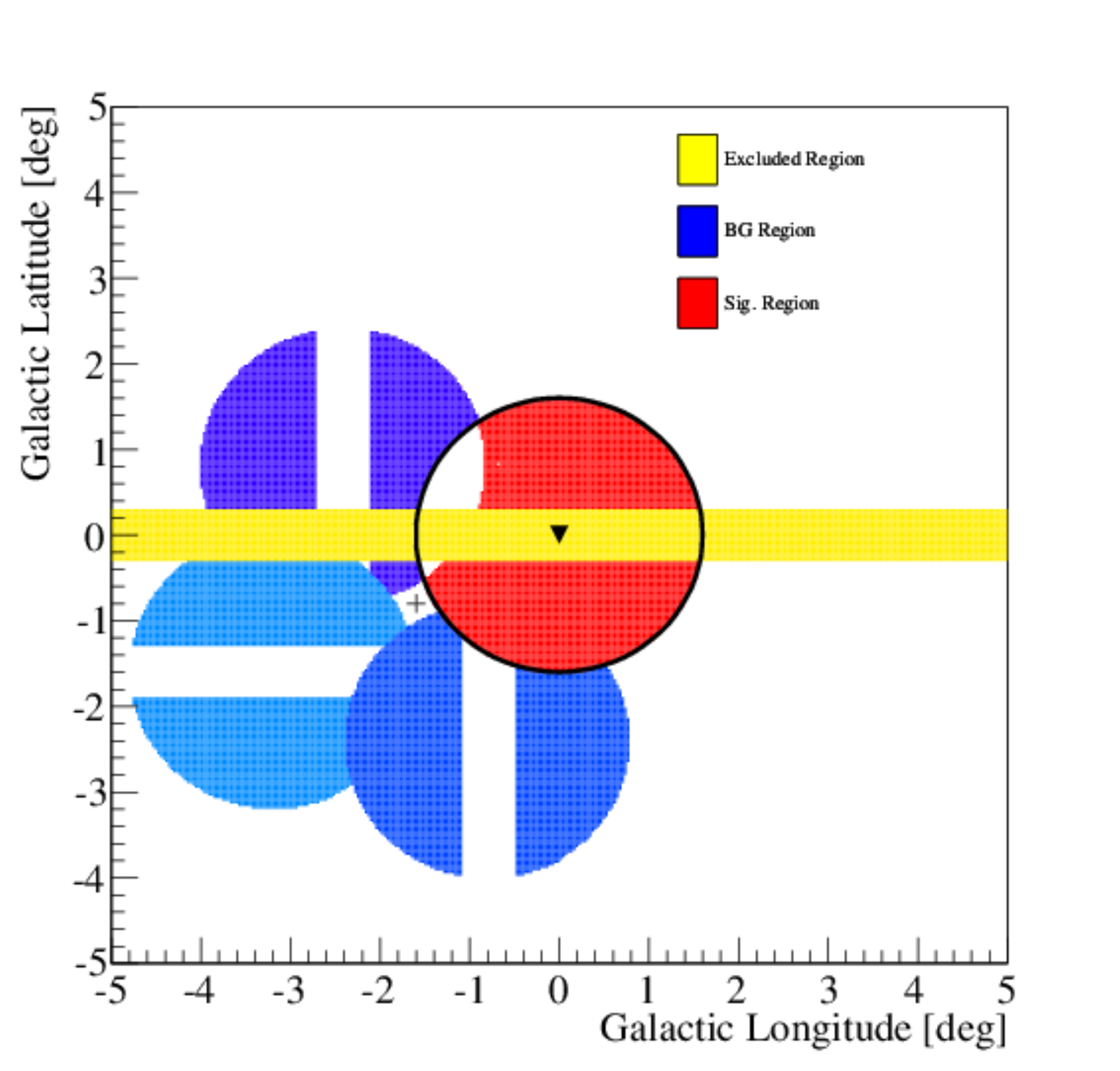}
  \caption{Diagram showing the reflected pixel technique as implemented for CTA. The Galactic Centre is marked with a black triangle, the observation position with a black cross. The signal region is coloured red, and is a circle around the Galactic Centre. The background regions are constructed by rotating pixels (segments) of the signal region about the observation position, ensuring equal acceptance by maintaining a constant angular distance from the observation position. The three blue regions are used as background. The galactic plane is excluded within a range of  $\pm0.3^{\circ}$, shown in yellow. White pixels within the signal region are those for which no background region was found.}
  \label{fig2}
 \end{figure}

Careful background subtraction techniques must be used to ensure that the large astrophysical contributions are properly treated. 
Two complementary background subtraction techniques are compared in this study, the Ring Background method~\cite{bib:apdm}, and the Reflected Pixel method~\cite{bib:gerretICRC}. These techniques are described in Figures~\ref{fig1} and ~\ref{fig2}.
\section{Simulations}
\subsection{CTA Performance}
\begin{figure}[htb]
  \centering
  \includegraphics[width=0.4\textwidth]{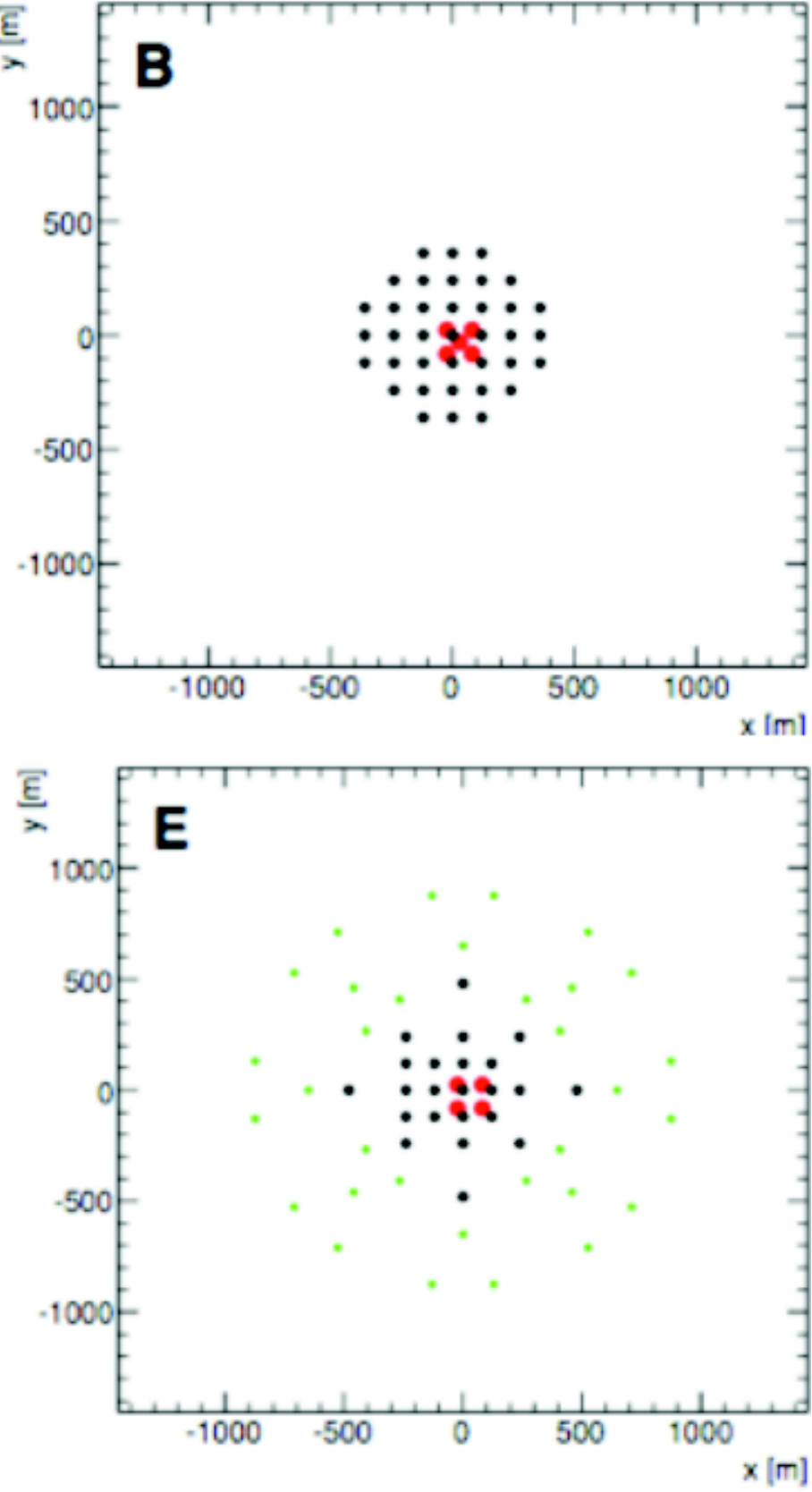}
  \caption{Two possible layouts for CTA. Array B (top) is a compact array consisting of MSTs (black) and LSTs (red); Array E (bottom) is more widely spread and also includes SSTs (green).  }
  \label{fig4}
 \end{figure}

\begin{figure}[htb]
  \centering
  \includegraphics[width=0.4\textwidth]{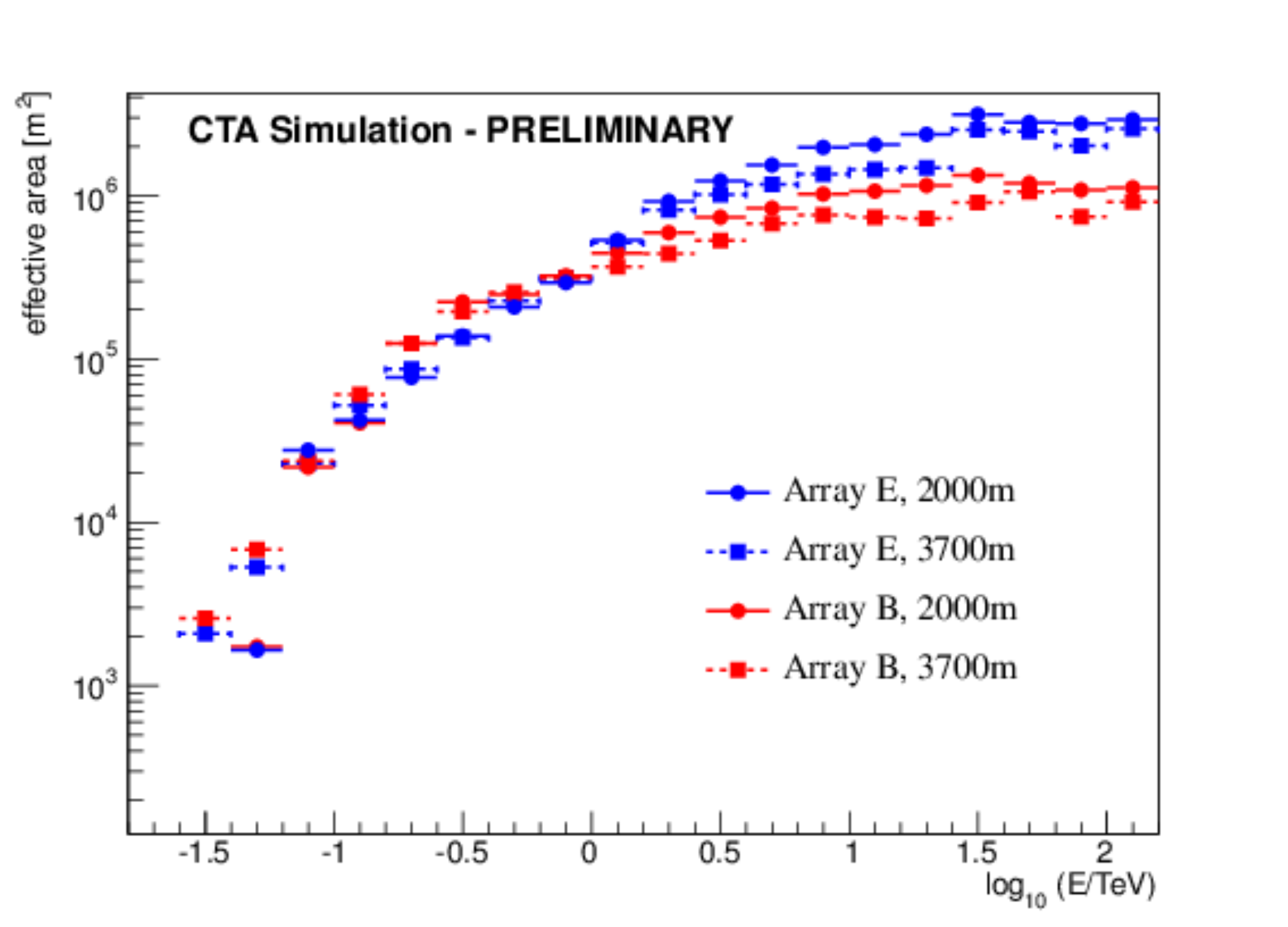}
  \caption{Energy dependent gamma-ray effective areas for two possible CTA array layouts (E and B) at altitudes of 2000m and 3700m, after gamma-hadron separation.    }
  \label{fig3}
 \end{figure}

The possible array layouts for CTA tested in this analysis are shown in Figure~\ref{fig4}.
Monte Carlo (MC) simulations of on-axis photons, electrons and protons, detected at 2000 m and 3700 m with a zenith angle of 20$^{\circ}$ are used to calculate the effective areas, as well as angular and energy resolutions for the potential arrays. For both altitudes, CORSIKA~\cite{bib:corsika} version 6.735 is used to simulate photons and electrons with an energy range of 0.003-300 TeV, and protons with 0.005-500 TeV, with a power law index of -2.  For the 2000 m case, a core scatter circle of 3000 m radius is assumed, for 3700 m altitude this radius is 2500 m. Figure~\ref{fig3} shows the effective areas as a function of photon energy after gamma-hadron separation for the two arrays and two altitudes under consideration. The effective areas shown are preliminary and the analysis, and telescope spacing within the arrays have not been optimised for the 3700 m case.  Although effective areas are expected to increase at higher altitudes for low energies, this is countered by an increase in the background rates. 
\subsection{Analysis simulations and assumptions}
In addition to the simulated performance of CTA described in terms of effective areas in the previous subsection, two further factors affect the sensitivity study: the background subtraction method (as discussed in Section~\ref{sec_bg}), and the characteristics of the DM distribution and annihilation in the Galactic Centre halo.

For the purpose of this study, the annihilating DM particles are assumed to be generic Weakly Interacting Massive Particles (WIMPs), with mass $m_{\chi}$. The expected rates in the signal and background regions (solid angles $\Delta\Omega_{s,b}$) are given by:
\begin{equation}
\frac{dR}{dE}|_{s,b} = \frac{<\sigma_{ann}v>}{8\pi m_{\chi}^2}\int_{\Delta\Omega_{s,b}} J(\Omega)A(\Omega,E)d\Omega,
\end{equation}
where $A(\Omega,E)$ is the effective area for photons (dependent on the position within the field of view), $\Omega_{s,b}$ is the solid angle of the signal or background region, over which the flux is considered, $<\sigma_{ann}v>$ is the velocity averaged annihilation cross section and $m_{\chi}$ is the WIMP mass.  The astrophysical factor, $J(\Omega)$ is the line of sight integral over the squared DM density, taken from N-body simulations~\cite{bib:Nbody}.

For the main analysis, a generic Tasitsiomi spectrum~\cite{bib:tasit} is assumed, dominantly annihilating into quark-antiquark pairs with hadronisation into $\pi^0$ mesons.

The DM search sensitivity depends on observation time; an observation time of 100h, or 10$\%$ of data taking time for one year, is assumed here.

\section{Results}
\begin{figure}[t!]
  \centering
  \includegraphics[width=0.4\textwidth]{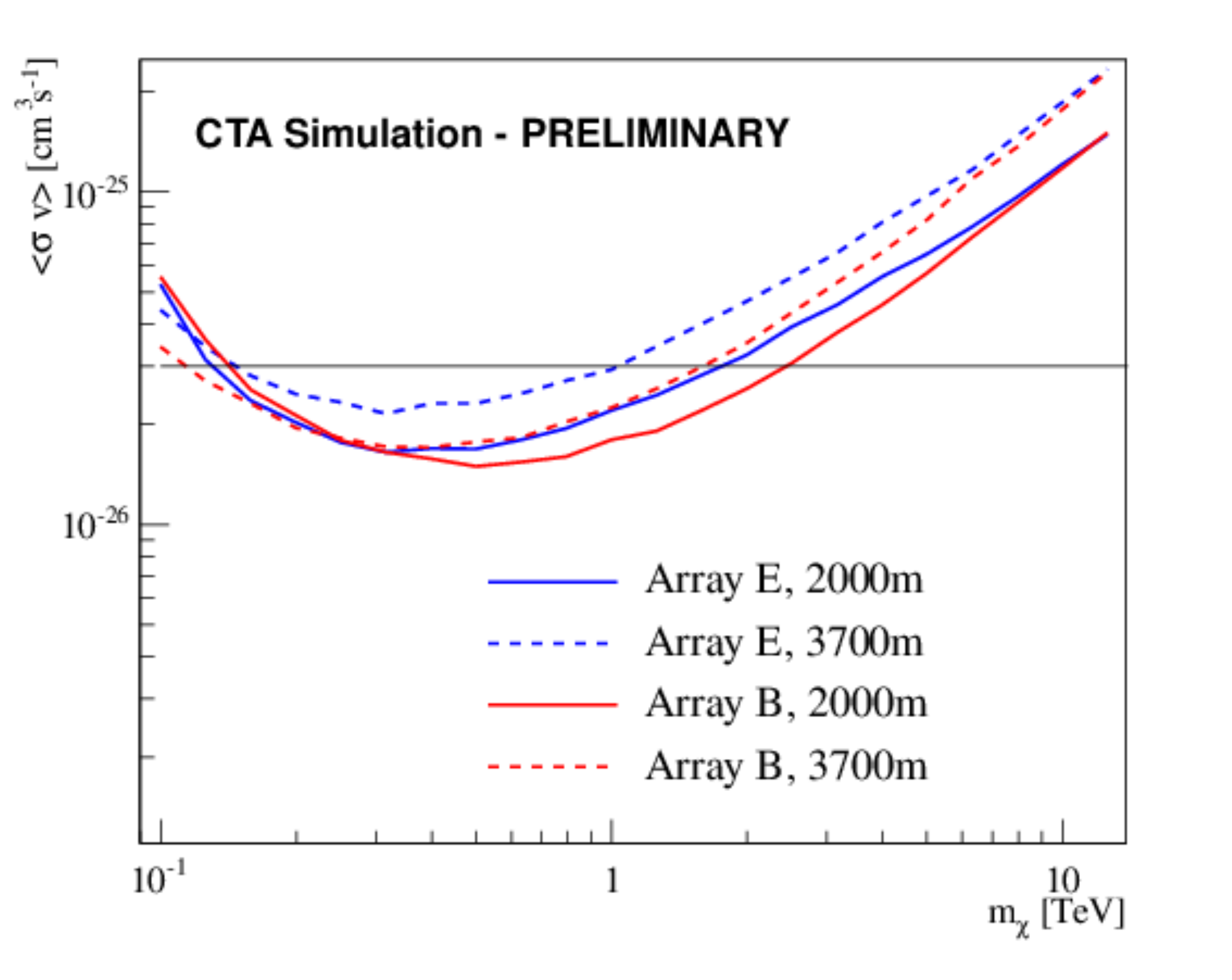}
  \includegraphics[width=0.4\textwidth]{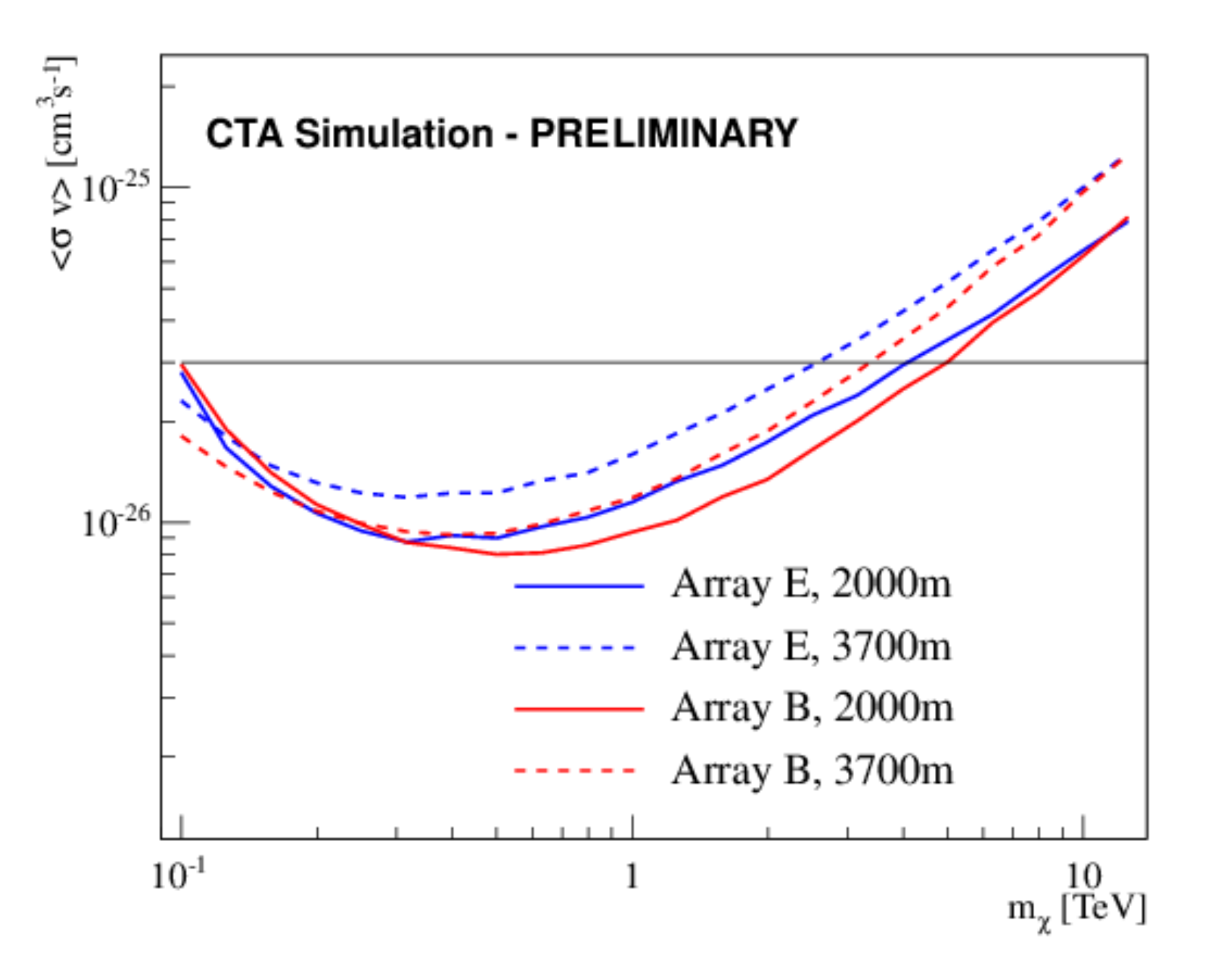}

  \caption{Sensitivity to the velocity averaged annihilation cross section as a function of WIMP mass for CTA arrays E (blue) and B (red). (Top) Ring background subtraction method, (Bottom) Reflected Pixel method. }
  \label{fig5}
 \end{figure}

The sensitivity of CTA to WIMP dark matter is considered under the various sets of assumptions described in the previous sections. Figure~\ref{fig5} shows the sensitivity curves in velocity averaged annihilation cross section as a function of WIMP mass, for the alternative background methods, altitudes and array layouts.  The DM sensitivity is shown for the two alternative background subtraction methods discussed; the difference in sensitivity between these two methods dominates over any differences due to array layout and altitude.


\section{Conclusions and Future}
The results of this study alone are insufficient to determine whether a site at 2000m or 3700m would be more suitable for CTA. 
From Figure~\ref{fig5} it can be seen that differences in background subtraction technique have a larger effect than changes of altitude or array layout on the DM search sensitivity.  

These simulations will be combined with other sensitivity studies and factors including climate/weather observations, site access and background light levels, to assess the advantages of each possible site location before the final CTA sites are chosen.

%
%
%
%
%
%
%
%
%
%
%
%
%
%
%
%
%

%
\vspace*{0.5cm}
\footnotesize{{\bf Acknowledgment:}{We gratefully acknowledge support from the agencies and organisations listed at the following URL: http://www.cta-observatory.org/?q=node/22}}

\end{document}